\documentclass[acus]{JAC2001}

\usepackage{graphicx}

\begin{document}
\title{PARTICLE PHYSICS AT FUTURE COLLIDERS}

\author{John Ellis, {\it Theoretical Physics Division, CERN, Geneva, 
Switzerland}}

\maketitle
  
\begin{abstract}

The search for physics beyond the Standard Model motivates new high-energy
accelerators, which will require high luminosities in order to produce
interesting new heavy particles. Using the Higgs boson and supersymmetry
as examples, we discuss the capabilities of the LHC and $e^+ e^-$ linear
colliders in the TeV and multi-TeV energy ranges to discover and study new
particles.

\end{abstract}
\begin{center}
{CERN-TH/2002-292 ~~~~~~~~~~ hep-ex/0210052}\\
{\it Talk presented at the 26th Advanced ICFA Workshop on 
Nanometre-Size Colliding Beams, Lausanne, September 
2002}\\
\end{center}
\section{The Need for Nanobeams}

The primary motivation for future colliders, and the only one likely to 
find favour with funding agencies, is to search for new physics beyond the 
Standard Model. This may be done either by going to higher energies, or by 
colliding beams with higher luminosities in the energy range already 
probed by previous colliders. In general, the cross sections for 
interesting new physics processes decrease at higher energies:
\begin{equation}
\sigma_{\rm interesting} \; \sim \; {1 \over E_{\rm CM}^2}.
\label{fall}
\end{equation}
The basic reason for this decrease with energy is that the interesting 
cross sections are those for point-like particles whose effective sizes 
are determined by their Compton wavelengths $R \sim 1/E_{\rm CM}$. 
Likewise, 
interesting new particles with masses $M_{\rm new}$ have production cross 
sections
\begin{equation}
\sigma_{\rm new} \; \sim \;  {1 \over M_{\rm new}^2}.
\label{new}
\end{equation}
A suitable standard of comparison for high-energy future colliders is 
provided by LEP, which reached a maximum luminosity ${\cal L} \sim 
10^{32}$~cm$^{-2}$s$^{-1}$ at $E_{\rm CM} \sim 200$~GeV. The 
pair-production of heavy particles such as the $W^\pm$ and $Z^0$ at
LEP~2 was 
not overly generous, so we assume that a new collider should provide a
similar number of events. In this case, its luminosity should increase 
as $E_{\rm CM}^2$ compared to LEP. 

Thus, the LHC with a luminosity ${\cal L} \sim 10^{34}$~cm$^{-2}$s$^{-1}$
will be able to produce pairs of new particles each weighing $\sim 1$~TeV.
Likewise, a linear $e^+ e^-$ collider operating at $E{\rm CM} \sim 5$~TeV,
such as CLIC, should be designed with a luminosity ${\cal L} \sim
10^{35}$~cm$^{-2}$s$^{-1}$. Such large luminosities are also required for
advanced `factories' at lower energies:  a rule of thumb is that an
$n$'th-generation factory should have a luminosity $\sim 10^n$ times
greater than the first collider to explore the same energy range. This is
why the present $B$ factories aim at ${\cal L} \sim
10^{34}$~cm$^{-2}$s$^{-1}$, and one talks about ${\cal L} \sim
10^{35}$~cm$^{-2}$s$^{-1}$ for the next generation, {\it et seq.}... 
Nanobeams will certainly be in great demand.

\section{The Standard Model of Particle Physics}

\begin{figure}[htb]
\centering
\includegraphics*[width=65mm]{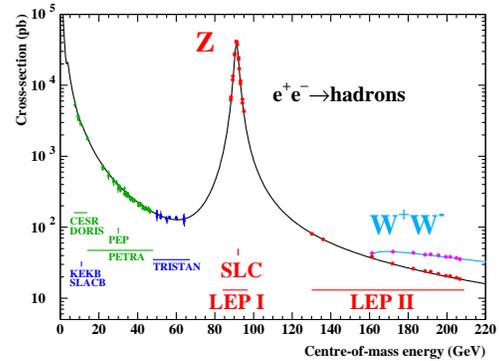}
\caption{Data from LEP and other $e^+ e^-$ experiments agree perfectly 
with the 
predictions of the Standard Model~\cite{Grunewald}.}
\label{fig:LEP}
\end{figure}

LEP and the lower-energy colliders that preceded it have established
beyond question the Standard Model of particle physics. It comprises three
generations of fundamental fermions to make up the matter in the Universe,
each consisting of two quarks, a neutrino and an electron-like charged
lepton. Four fundamental forces act on these matter particles: the
electromagnetic, strong, weak and gravitational forces.  Each of these is
carried by messenger particles: the photon, the gluons, the $W^\pm$ and
$Z^0$, and (we believe) the graviton, respectively. As seen in
Fig.~\ref{fig:LEP}, the experimental data from LEP agree (too)  
perfectly with the theoretical curves, at all energies up to above
200~GeV~\cite{Grunewald}. This sounds great, but there are plenty of
questions left open by the Standard Model.

Why are some fundamental particles, such as the photon and gluons,
massless, while others are massive, weighing as much as good-sized nuclei
in the cases of the $W^\pm$, $Z^0$ and top quark? Are the different
fundamental forces unified, as long hoped by Einstein? Why are there so
many different types of `elementary' particles? Could they all be
composite states made out of more fundamental constituents? How to explain
all the different parameters of the Standard Model: 6 quark masses, 3
charged-lepton masses, 2 weak-boson masses, 4 weak mixing angles and
phases, 3 interaction strengths and a non-perturbative strong-interaction
vacuum parameter?  These total 19 parameters, even without describing
neutrino masses and mixing angles.

Some more fundamental physics must surely lie beyond the Standard Model, 
and the next sections of this paper describe some candidates for this new 
physics.

\section{The Problem of Mass}

This is probably the most pressing problem raised by the Standard Model.
Indeed, it can only be solved by introducing new physics at some energy
scale below $\sim 1$~TeV. The most likely culprit for generating particle
masses is thought to be a Higgs boson with a mass in this range.
A massless vector particle such as the photon has two polarization
states:  $\lambda = -1, +1$. On the other hand, a massive vector particle
such as the $W^\pm$ or $Z^0$ must have three polarization states: $\lambda
= -1, 0, +1$. Thus, in order for a massless vector particle to acquire a
mass, it must combine with some zero-polarization state, such as could be
provided by a spin-0 field via the Higgs-Brout-Englert 
mechanism~\cite{HBE}.

In the Standard Model, the minimal such model contains a complex doublet
of Higgs fields, with a total of four degrees of freedom. Of these, three 
are
eaten by the $W^\pm$ and $Z^0$ to become their third polarization states,
leaving one degree of freedom to appear as a separate physical state, the
Higgs boson. In order for it to perform its task of giving masses to other
particles, its couplings to them should be proportional to their masses:  
$g_{H {\bar f} f} \propto m_f$. However, the mass of the physical Higgs
boson itself is not fixed in the Standard model without any extra input.

\begin{figure}[htb]
\centering
\includegraphics*[width=65mm]{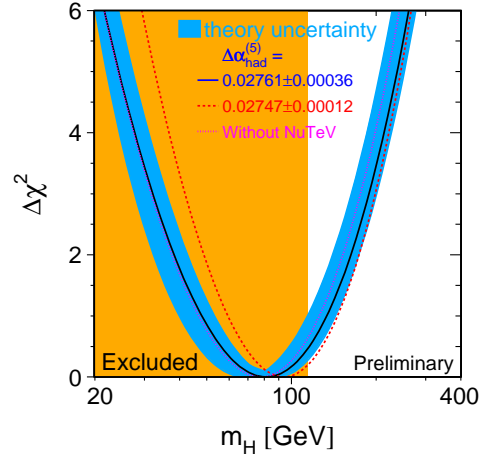}
\caption{The $\chi^2$ curve for a global fit to the precision electroweak 
data from LEP and elsewhere, with the uncertainties shaded in 
blue~\cite{Grunewald}, favour a relatively light Higgs boson with mass 
close to 
the range excluded by experiment, shaded in yellow~\cite{LEPHWG}.}
\label{fig:blue}
\end{figure}

\begin{figure}[htb]
\centering
\includegraphics*[width=65mm]{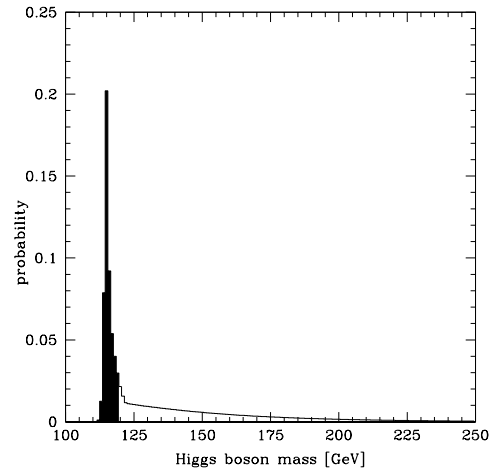}
\caption{An estimated probability distribution for the Higgs 
mass~\cite{Erler}, obtained by 
convoluting the blue-band plot in Fig.~\ref{fig:blue}~\cite{Grunewald} 
with the experimental exclusion~\cite{LEPHWG}.}
\label{fig:Erler}
\end{figure}

Direct searches for the Higgs boson at LEP have established that the Higgs
boson weighs more than $114.4$~GeV~\cite{LEPHWG}. Precision electroweak
data from LEP and elsewhere also provide indirect information on the
possible mass of the Higgs boson, as seen in
Fig.~\ref{fig:blue}~\cite{Grunewald}. Quantum corrections in the Standard
Model would disagree with the precision measurements unless the Higgs
boson weighs less than 193~GeV at the 95~\% confidence level, with a mass
$\sim 115$~GeV being the most likely value, as seen in
Fig.~\ref{fig:Erler}~\cite{Erler}. This probability distribution makes no
use of the `hint' from direct Higgs searches at LEP of a signal at $\sim
116$~GeV~\cite{LEPHWG}.

\begin{figure}[htb]
\centering
\includegraphics*[width=65mm]{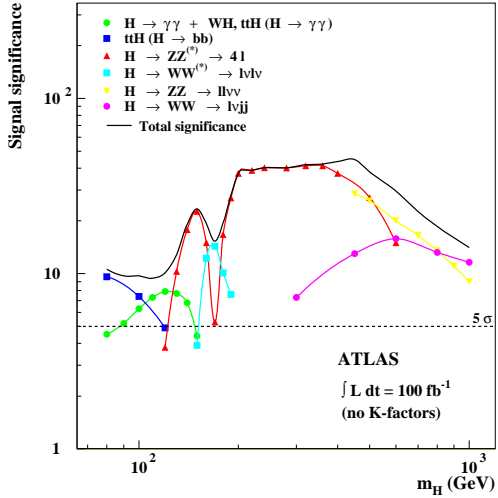}
\caption{The LHC experiments will be able to discover the Higgs boson with 
high significance, whatever its mass, and may observe several 
of its decay modes~\cite{LHC}.}
\label{fig:LHC}
\end{figure}

Now that LEP operations have been terminated, what are the prospects for
Higgs searches with future colliders? The Tevatron collider has a chance,
if it can accumulate sufficient luminosity, particularly if the Higgs
boson weighs $\sim 115$~GeV~\cite{Tevatron}. The LHC will be able to
discover the Higgs boson, whatever it mass below about 1~TeV, as well as
observe two or three of its decay modes and measure its mass to 1\% or
better, as seen in Fig.~\ref{fig:LHC}~\cite{LHC}. The days of the Higgs
boson are numbered!

\section{Supersymmetry}

The Higgs boson is confidently expected even within the Standard Model,
but theorists think it should also be accompanied by some new physics
beyond the Standard Model. The reason for this is to help understand the
hierarchy of different mass scales in physics, and in particular why $m_W
\ll m_P \sim 10^{19}$~GeV, the Planck mass scale where gravity is expected
to become strong and the only candidate we have for a fundamental mass
scale in physics. Equivalently, we might ask why there is a hierarchy of
different interaction strengths: $G_F \sim 1/m_W^2 \gg G_N = 1/m_P^2$, or
why the Coulomb potential $\sim 1 /r$ inside an atom is so much larger
than the Newton potential $\sim G_N m^2 = (m / m_P)^2$.

You might think one could just `set and forget' the mass hierarchy, but 
then it would be upset by quantum corrections. Typical one-loop diagrams 
in the Standard Model make contributions to the Higgs and 
$W^\pm$ masses that diverge quadratically:
\begin{equation}
\delta m_{H, W}^2 \; \sim \; {\cal O}({\alpha \over \pi}) \Lambda^2,
\label{quadratic}
\end{equation}
where $\Lambda$ is a cutoff representing the energy scale at which new 
physics should appear. The quantum `correction' ({\ref{quadratic}) would 
be 
much larger than the physical value of $m_W$ if the new physics scale 
$\Lambda \sim m_P$. However, the `correction' (\ref{quadratic}) could be 
made naturally small by postulating equal numbers of bosons $B$ and 
fermions $F$
(whose loop diagrams have opposite signs) with equal coupling strengths 
$\alpha$. In this case, (\ref{quadratic}) would be replaced by
\begin{equation}
\delta m_{W, H}^2 \; \sim \; {\cal O}({\alpha \over \pi}) \left( m_B^2 - 
m_F^2 \right),
\label{susy}
\end{equation}
which would be comparable to $m_{W, H}^2$ if
\begin{equation}
\vert m_B^2 - m_F^2 \vert \; \sim \; 1~{\rm TeV}^2.
\label{TeV}
\end{equation}
This is the motivation for low-energy supersymmetry~\cite{hierarchy}.

There is no direct evidence for supersymmetry, but there are several
indirect hints that supersymmetry may indeed appear at some energy scale
below about 1~TeV. One is provided by the strengths of the
electromagnetic, weak and strong interactions measured at LEP, which do
not extrapolate to a common unified value in the absence of supersymmetry,
but do unify at high energies if supersymmetric particles wighing $\sim
1$~TeV are included in the renormalization-group equations~\cite{GUT}, as 
seen in
Fig.~\ref{fig:GUT}. Another hint is provided by the likely mass of the
Higgs boson. In models with low-energy supersymmetry, it is calculated to
weigh less than about 130~GeV~\cite{susyHiggs}, highly consistent with the 
range suggested
by the precision electroweak data. A third hint may be provided by the
dark matter thought to abound in the Universe. The lightest supersymmetric
particle (LSP) is stable in the minimal supersymmetric extension of the
Standard Model (MSSM), and would be an ideal particle candidate for dark
matter if it weighs less than about 1~TeV~\cite{EHNOS}.

\begin{figure}[htb]
\begin{flushleft}
\includegraphics*[width=150mm]{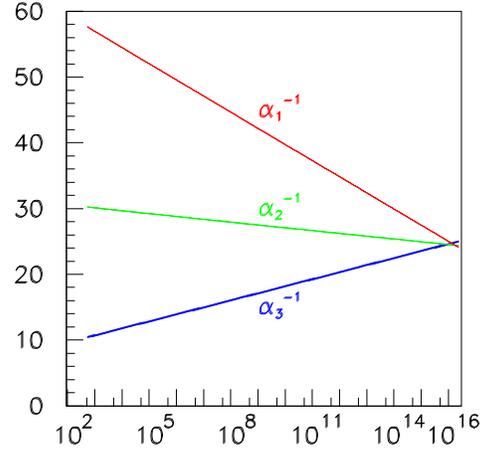}
\end{flushleft}
\vspace{-8cm}
\caption{The measurements (vertical axis) of the gauge coupling strengths 
of the 
Standard Model at LEP and elsewhere can be evolved up to high energies 
(horizontal axis, in units of GeV)
using renormalization-group equations incorporating supersymmetry. They 
are consistent with unification at a very high energy scale, but not with 
unification without supersymmetry~\cite{GUT}.}
\label{fig:GUT}
\end{figure}

On the other hand, no sparticles have ever been seen, in particular at
LEP, imposing important constraints on the MSSM~\cite{LEPSWG}. For
example, charginos - the supersymmetric partners of the $W^\pm$ - must
weigh more than about 103~GeV, and slectrons - the supersymmetric partners
of the electron - must weigh more than about 100~GeV. The lower limit on
the mass of the Higgs boson, mentioned above, also imposes an important
constraint on the MSSM parameter space, as does the agreement between
Standard Model calculations and the experimental rate of $b \to s \gamma$
decay, as seen in Fig.~\ref{fig:CMSSM}~\cite{EFOS}.

\begin{figure}[htb]
\centering
\includegraphics*[width=65mm]{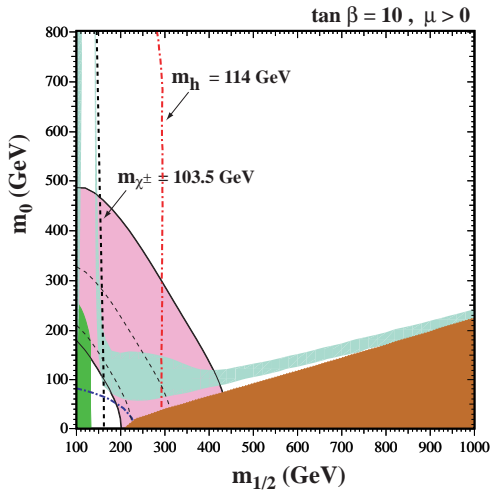}
\caption{The parameter space of the MSSM projected onto the $(m_{1/2}, m_0)$ 
plane for $\tan \beta = 10$ and $\mu > 0$. The LEP lower limits on the 
Higgs, chargino and selectron masses are shown as (red) dot-dashed, 
(black) dashed and (blue) dash-dotted lines, respectively. The region 
at small $(m_{1/2}, m_0)$ excluded by $b \to s \gamma$ is shaded 
(green). The dark (red) shaded 
region is excluded because dark matter must be neutral, and the region 
where its relic density falls within the range preferred by cosmology has 
light (turquoise) shading. The region preferred by the BNL measurement of 
$g_\mu -2$ and low-energy $e^+e^-$ data is shaded (pink)~\cite{EFOS}.} 
\label{fig:CMSSM} 
\end{figure}

As also seen in Fig.~\ref{fig:CMSSM}, a further experimental constraint is
provided by the recent measurement of the anomalous magnetic moment of the
muon, $g_\mu - 2$, even if it does not disagree significantly with the
Standard Model~\cite{g-2}. As things stand, the measured value of $g_\mu -
2$ disagrees by 3 standard deviations with the best estimate based on $e^+
e^- \to$~hadrons data~\cite{threesigma}, though the discrepancy with
estimates based on $\tau \to$~hadrons data is less than 2 standard
deviations.

\section{Benchmark Supersymmetric Scenarios}

As seen in Fig.~\ref{fig:CMSSM}, all these constraints on the MSSM are
mutually compatible. As an aid to understanding better the physics
capabilities of the LHC, various linear $e^+e^-$ linear collider designs
and non-accelerator experiments, a set of benchmark supersymmetric
scenarios have been proposed~\cite{Bench}. These are compatible with all
the accelerator constraints mentioned above, including the LEP searches
and $b \to s \gamma$, and yield relic densities of LSPs in the range
suggested by cosmology and astrophysics. These benchmarks are not
inteneded to sample `fairly' the allowed parameter space, but rather to
illustrate the range of possibilities currently allowed, as shown in
Fig.~\ref{fig:bench}.

\begin{figure}[htb]
\centering
\includegraphics*[width=65mm]{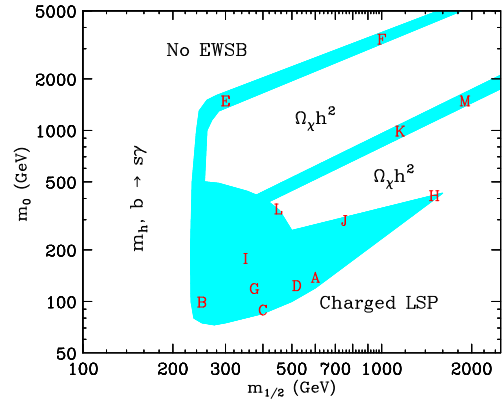}
\caption{Sketch of the distribution of proposed CMSSM benchmark points
in the $(m_{1/2}, m_0)$ plane~\cite{Bench}. These points were chosen as 
illustrations of the range of possibilities in the CMSSM, rather than as 
a `fair' sample of its parameter space.} 
\label{fig:bench}
\end{figure}

In addition to a number of benchmark points falling in the `bulk' region
of parameter space at relatively low values of the supersymmetric particle
masses, we also proposed some points out along the `tails' of parameter
space extending out to larger masses. These clearly require some degree of
fine-tuning to obtain the required relic density~\cite{EO} and/or the
correct $W^\pm$ mass~\cite{EOS}, and some are also disfavoured by the
supersymmetric interpretation of the $g_\mu - 2$ anomaly, but all are
logically consistent possibilities.

\section{LHC Physics}

The cross sections for producing pairs of supersymmetric particles at the
LHC decrease with increasing masses. Nevertheless, the signature expected
for supersymmetry - multiple jets and/or leptons with a large amount of
missing energy - is quite distinctive. Therefore, the detection of the
supersymmetric partners of quarks and gluons at the LHC is expected to be
quite easy if they weigh less than about 2~TeV~\cite{LHC}. Moreover,
in many scenarios one should be able to observe their cascade decays into
lighter supersymmetric particles~\cite{Paige}. As seen in
Fig.~\ref{fig:Manhattan}, large fractions of the supersymmetric spectrum
should be seen in most of the benchmark scenarios, although there are a
couple where only the lightest supersymmetric Higgs boson would be
seen~\cite{Bench}.

\begin{figure}[htb]
\centering
\includegraphics*[width=65mm]{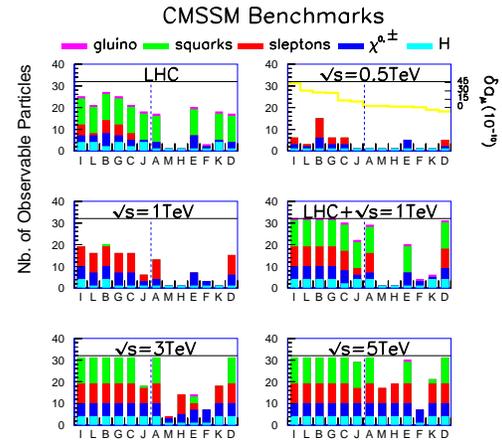}
\caption{Estimates of the numbers of different CMSSM particles that could 
be seen in each of the benchmark scenarios~\cite{Bench}, at each of the 
indicated colliders. The points are ordered from left to right by their 
decreasing compatibility with the BNL $g_\mu - 2$ measurement.} 
\label{fig:Manhattan} 
\end{figure}

\section{$e^+ e^-$ Linear Collider Physics}

Electron-positron colliders provide very clean experimental environments, 
with egalitarian production of all the new particles that are 
kinematically accessible, including those that have only weak 
interactions. Moreover, polarized beams provide a useful analysis tool, 
and $e \gamma$, $\gamma \gamma$ and $e^- e^-$ colliders are readily 
available at relatively low marginal costs.

The $e^+ e^- \to {\bar t} t$ threshold is known to be at $E_{\rm CM} \sim
350$~GeV. Moreover, if the Higgs boson indeed weighs less than 200~GeV, as
suggested by the precision electroweak data, its production and study
would also be easy at an $e^+ e^-$ collider with $E_{\rm CM} \sim
500$~GeV. With a luminosity of $10^{34}$~cm$^{-2}$s$^{-1}$ or more, many
decay modes of the Higgs boson could be measured very
accurately~\cite{TESLA}, and one might be able to find a hint whether its
properties were modified by supersymmetry~\cite{EHOW}.

However, the direct production of supersymmetric particles at such a
collider cannot be guaranteed~\cite{EFGO}, as seen in
Fig.~\ref{fig:Manhattan}~\cite{Bench}. We do not yet know what the
supersymmetric threshold energy may be (or even if there is one!). We may
well still not know before the operation of the LHC, although $g_\mu - 2$
might provide an indication, if the uncertainties in the Standard Model
calculation can be reduced.

If an $e^+ e^-$ collider is above the supersymmetric threshold, it will be
able to measure very accurately the sparticle masses. By comparing their
masses with those of different sparticles produced at the LHC as seen in
Fig.~\ref{fig:BPZ}, one would be able to make interesting tests of string
and GUT models of supersymmetry breaking~\cite{Bench}. However,
independently from the particular benchmark scenarios proposed, a linear
$e^+ e^-$ collider with $E_{\rm CM} < 1$~TeV would not cover all the
supersymmetric parameter space allowed by cosmology.

\begin{figure}[htb]
\begin{flushleft}
\includegraphics*[width=160mm]{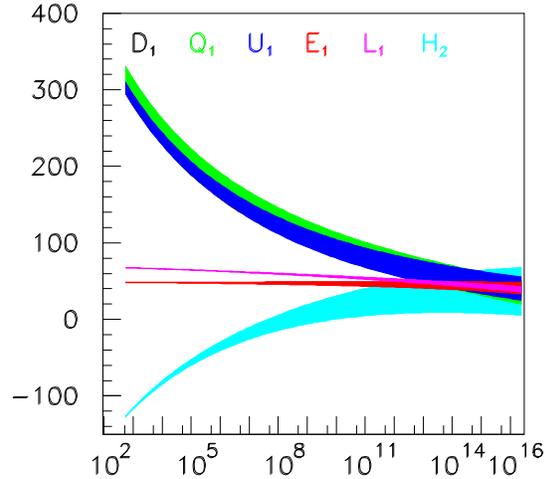}
\end{flushleft}
\vspace{-8cm}
\caption{Analogously to the unification of the gauge couplings shown in 
Fig.~\ref{fig:GUT}, measurements of the sparticle masses at future 
colliders (vertical axis, in units of GeV) can be evolved up to high 
scales (horizontal axis, in units of GeV) to test models of supersymmetry 
breaking, in particular whether squark and slepton masses are universal at 
some input GUT scale~\cite{BPZ}.}
\label{fig:BPZ}
\end{figure}

Nevertheless, there are compelling physics arguments for such a linear 
$e^+ e^-$ collider, which would be very complementary to the LHC in terms 
of its exploratory power and precision. It is to be hoped that the world 
community will converge on a single project with the widest possible 
energy range.

\section{CLIC}

CERN and its collaborating institutes are studying the possible
following step in linear $e^+ e^-$ colliders, a multi-TeV machine called
CLIC~\cite{CLIC}. This would use a double-beam technique to attain
accelerating gradients as high as 150~MV/m, and the viability of
accelerating structures capable of achieving this field has been
demonstrated in the CLIC test facility~\cite{CTF}. Parameter sets have
been calculated for CLIC designs with $E_{\rm CM} = 3$ and $5$~TeV, and
luminosities of $10^{35}$~cm$^{-2}$s$^{-1}$ or more.

In many of the proposed benchmark supersymmetric scenarios, CLIC would be 
able to complete the supersymmetric spectrum and/or measure in much more 
detail heavy sparticles found previously at the LHC. CLIC produces more 
beamstrahlung than lower-energy linear $e^+ e^-$ colliders, but the 
supersymmetric missing-energy signature would still be easy to 
distinguish, and accurate measurements of masses and decay modes could 
still be made, as seen in Fig.~\ref{fig:susyCLIC}~\cite{Battaglia}.

\begin{figure}[htb]
\centering
\includegraphics*[width=65mm]{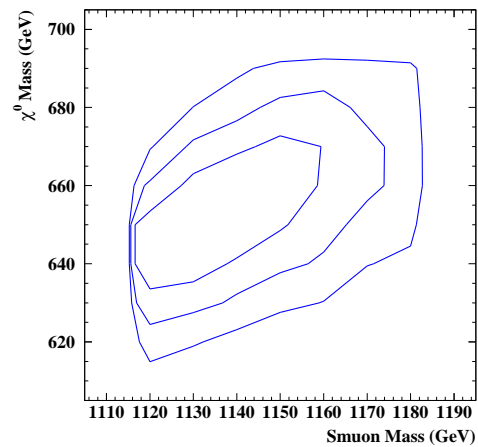}
\caption{Like lower-energy $e^+ e^-$ colliders, CLIC enables very accurate 
measurements of sparticle masses to be made, in this case the 
supersymmetric partner of the muon and the lightest 
neutralino $\chi^0$~\cite{Battaglia}.}
\label{fig:susyCLIC}
\end{figure}

\section{Perspectives}

In this brief talk, I have tried to explain why higher-energy physics 
requires higher luminosities and hence smaller beams. I have used the 
Higgs boson and supersymmetry as examples of the new physics that may be 
awaiting us at the TeV scale, and shown how they could be explored by 
colliders with luminosities that are sufficiently high. Other examples, 
including extra dimensions, are considered in~\cite{DEG}. One can already 
say that linear $e^+ e^-$ colliders with energies in the sub- and 
multi-TeV ranges would both be interesting.

What ideas exist for colliders to achieve even higher energies? One 
possibility might be a VLHC with $E_{\rm CM} \sim 100$~TeV or more. In 
order to realize its full kinematic potential, such a machine should have 
a luminosity of $10^{35}$~cm$^{-2}$s$^{-1}$ or more, favouring very small 
beams. As for leptons, $e^+ e^-$ colliders with $E_{\rm CM} \sim 10$~TeV 
or more are very difficult to imagine. An alternative might be a very 
high-energy $\mu^+ \mu^-$ collider, but this would have to surmount the 
hurdles of muon cooling and neutrino radiation.

Even before such futuristic devices, there will be plenty of work for the 
nanobeam community.


\begin{thebibliography}{99}

\bibitem{Grunewald}
M.~W.~Grunewald,
``Electroweak physics",
arXiv:hep-ex/0210003.

\bibitem{HBE}
P.~W.~Higgs,
Phys.\ Lett.\  {\bf 12} (1964) 132 and
Phys.\ Rev.\ Lett.\  {\bf 13} (1964) 508;
F.~Englert and R.~Brout,
Phys.\ Rev.\ Lett.\  {\bf 13} (1964) 321.

\bibitem{LEPHWG}
LEP Higgs Working Group, \\
{\tt http://lephiggs.web.cern.ch/LEPHIGGS/papers/}
{\tt July2002{\_}SM/index.html}.

\bibitem{Erler}
J.~Erler,
Phys.\ Rev.\ D {\bf 63} (2001) 071301
[arXiv:hep-ph/0010153].

\bibitem{Tevatron}
M.~Carena {\it et al.},
``Report of the Tevatron Higgs working group",
arXiv:hep-ph/0010338.

\bibitem{LHC}
ATLAS Collaboration, \\
{\tt http://atlas.web.cern.ch/Atlas/GROUPS/}
{\tt PHYSICS/TDR/access.html}; \\
CMS Collaboration, \\
{\tt 
http://cmsdoc.cern.ch/cms/outreach/html/}
{\tt CMSdocuments/CMSplots/CMSplots.html}.

\bibitem{hierarchy}
L.~Maiani, {\it Proceedings of the 1979 Gif-sur-Yvette Summer School On
Particle
Physics}, 1;
G.~'t Hooft, in {\it Recent Developments in Gauge Theories, Proceedings
of the Nato Advanced Study
Institute, Cargese, 1979}, eds. G.~'t Hooft {\it et al.}, (Plenum Press,
NY, 1980); E.~Witten,
Phys.\ Lett.\  B {\bf 105} (1981) 267.

\bibitem{GUT}
J.~Ellis, S.~Kelley and D.~V.~Nanopoulos,
Phys.\ Lett.\ B {\bf 260} (1991) 131;
U.~Amaldi, W.~de Boer and H.~Furstenau,
Phys.\ Lett.\ B {\bf 260} (1991) 447;
C.~Giunti, C.~W.~Kim and U.~W.~Lee,
Mod.\ Phys.\ Lett.\ A {\bf 6} (1991) 1745.

\bibitem{susyHiggs}
J.  Ellis, G.  Ridolfi and F.  Zwirner, Phys.\ Lett.\
B {\bf 257} (1991) 83; M.S.  Berger, Phys.\ Rev.\ D {\bf 41} (1990) 225;
Y.  Okada, M.  Yamaguchi and T. Yanagida, Prog.\ Theor.\ Phys.\ {\bf
85} (1991) 1; Phys.\ Lett.\ B {\bf 262} (1991) 54; H.E.  Haber and R.
Hempfling, Phys.\ Rev.\ Lett.\ {\bf 66} (1991) 1815; for the most recent 
calculations, see: S.~Heinemeyer, W.~Hollik and G.~Weiglein,
             {\em Comput. Phys. Commun.} {\bf 124} (2000) 76,
             hep-ph/9812320;
             hep-ph/0002213.

\bibitem{EHNOS}
J. Ellis, J.S. Hagelin, D.V. Nanopoulos, K.A. Olive
and M. Srednicki, Nucl. Phys. B {\bf 238} (1984) 453; see also
H. Goldberg, Phys. Rev. Lett. {\bf 50} (1983) 1419.

\bibitem{LEPSWG}
LEP Supersymmetry Working Group, \\
{\tt http://lepsusy.web.cern.ch/lepsusy/}.

\bibitem{EFOS}
J.~Ellis, T.~Falk, K.~A.~Olive and Y.~Santoso,
arXiv:hep-ph/0210205.

\bibitem{g-2}
G.~W.~Bennett {\it et al.}  [Muon g-2 Collaboration],
Phys.\ Rev.\ Lett.\  {\bf 89} (2002) 101804
[Erratum-ibid.\  {\bf 89} (2002) 129903]
[arXiv:hep-ex/0208001].

\bibitem{threesigma}
M.~Davier, S.~Eidelman, A.~Hocker and Z.~Zhang,
                 hep-ph/0208177; see also 
                 K.~Hagiwara, A.~Martin, D.~Nomura and T.~Teubner,
                 hep-ph/0209187; 
                 F.~Jegerlehner, unpublished, as reported in
                 M.~Krawczyk,
                 hep-ph/0208076.

\bibitem{Bench}
M.~Battaglia {\it et al.},
Eur.\ Phys.\ J.\ C {\bf 22} (2001) 535
[arXiv:hep-ph/0106204].

\bibitem{EO}
J.~R.~Ellis and K.~A.~Olive,
Phys.\ Lett.\ B {\bf 514} (2001) 114
[arXiv:hep-ph/0105004].

\bibitem{EOS}
J.~R.~Ellis, K.~A.~Olive and Y.~Santoso,
New J.\ Phys.\  {\bf 4} (2002) 32
[arXiv:hep-ph/0202110].

\bibitem{Paige}
I.~Hinchliffe, F.~E.~Paige, M.~D.~Shapiro, J.~Soderqvist and W.~Yao,
Phys.\ Rev.\ D {\bf 55} (1997) 5520
[arXiv:hep-ph/9610544].

\bibitem{TESLA}
TESLA TDR Part~3: {\it Physics at an $e^+e^-$
                   Linear Collider},
                   eds. R.D.~Heuer, D.~Miller, F.~Richard and P.M.~Zerwas,
                   hep-ph/0106315,
                   see: {\tt http://tesla.desy.de/tdr}.

\bibitem{EHOW}
J.~R.~Ellis, S.~Heinemeyer, K.~A.~Olive and G.~Weiglein, 
CERN--TH/2002-289.

\bibitem{EFGO}
J.~R.~Ellis, G.~Ganis and K.~A.~Olive,
Phys.\ Lett.\ B {\bf 474} (2000) 314
[arXiv:hep-ph/9912324].

\bibitem{BPZ}
G.~A.~Blair, W.~Porod and P.~M.~Zerwas,
arXiv:hep-ph/0210058.

\bibitem{CLIC}
R.~W.~Assmann {\it et al.},
``A 3-TeV e+ e- linear collider based on CLIC technology",
CERN report 2000-008; see also the CLIC web page: \\
{\tt http://cern.web.cern.ch/CERN/Divisions/PS/}
{\tt CLIC/Welcome.html}.

\bibitem{CTF}
H.~H.~Braun, S.~Dobert, I.~Syrachev, M.~Taborelli, I.~Wilson and 
W.~Wuensch, ``CLIC high-gradient test results",
CERN-PS-2002-062-RF.

\bibitem{Battaglia}
M. Battaglia, private communication.

\bibitem{DEG}
A.~De Roeck, J.~R.~Ellis and F.~Gianotti,
``Physics motivations for future CERN accelerators",
arXiv:hep-ex/0112004.

\end{thebibliography}
\end{document}